\documentclass[12pt,preprint]{aastex}
\shortauthors{Jiang et al.}

\begin{document}

\title{Weighing Super-Massive Black Holes with Narrow Fe K$\alpha$ Line}
\author{Peng Jiang, Junxian Wang and Xinwen Shu}
\affil{CAS Key Laboratory for Research in Galaxies and Cosmology, Department of Astronomy,
University of Science and Technology of China, Hefei, Anhui 230026, China}
\email{jpaty@mail.ustc.edu.cn}

\begin{abstract}
It has been suggested that the narrow cores of the Fe K$\alpha$ emission lines
in Active Galactic Nuclei (AGNs) are likely produced in the torus, the inner 
radius of which can be measured by observing the lag time between the 
$V$ and $K$ band flux variations. 
In this paper, we compare the virial products of the infrared time lags, and 
the narrow Fe K$\alpha$ widths for 10 type 1 AGNs, with the black hole masses from other 
techniques. We found the narrow Fe K$\alpha$ line width is in average 
2.6$^{+0.9}_{-0.4}$ times broader than expected, assuming an isotropic velocity 
distribution of the torus at the distance measured by the infrared lags.
We propose the thick disk model of the torus may explain the observed
larger line width. Another possibility is the contamination
by emission from the broad line region or the outer accretion disk.
Alternatively, the narrow iron line might originate from the inner most
part of the obscuring torus within the sublimation radius,
while the infrared emission may be from the outer cooler part.
We note the correlations between the black hole masses based
on this new technique and those based on other known techniques
are statistically insignificant. We argue that this 
could be attributed to the small sample size and the very large
uncertainties in the measurements of iron K line widths.
The next generation of X-ray observatories could help verify the origin of the
narrow iron K$\alpha$ line and the
reliability of this new technique.
\end{abstract}

\keywords{black hole physics -- galaxies: active -- galaxies: nuclei -- galaxies: Seyfert -- X-rays: galaxies -- infrared: galaxies}


\section{INTRODUCTION}
Under the unified scheme of Active Galactic Nuclei (AGNs),
Seyfert 1 nuclei with broad line regions (BLRs) would
be classified as Seyfert 2 if the BLRs were obscured
by encircling dust tori [1]. As nearly all of the
parsec-sized dust tori can not be spatially resolved
(note that Jaffe et al. reported interferometric mid-infrared
observations that spatially resolve the dust torus in NGC 1068 [2]),
the size and geometry of torus are not well understood. 
Recent researches measured the inner radius of dust torus by observing
the time-delayed responses of the $K$ band flux variations
to the $V$ band flux variations, as the bulk of the $K$ band flux should 
originate in the
thermal radiation of hot dust surrounding the central engine [3,4,5].

The iron K$\alpha$ emission line at $\sim$ 6.4 keV was first identified as
a common feature in the X-ray spectrum of AGNs by
{\it Ginga} [6,7].
The line can be promptly interpreted as fluorescence emission following
photoelectric absorption of the hard X-ray continuum [8].
Recent {\it XMM-Newton} and {\it Chandra}
observations revealed narrow (unresolved by {\it XMM}) iron K$\alpha$ lines
at $\sim$ 6.4 keV in the X-ray spectra of most AGNs [9,10,11]
However, the origin of the narrow iron K$\alpha$ line is still poorly
understood. Possible origins of the narrow line include the outermost regions
of the accretion disk, the BLR, and the dust torus.
Nandra found that the average Fe K$\alpha$ core emission width in
a sample of type 1 AGNs is about a factor of two narrower than the broad
emission line width (specifically H$\beta$), and there is no correlation 
between them [12]. This suggests that the iron K$\alpha$ emission
lines are likely, in many cases, originated in the torus at 
larger scales but not the BLRs.

Reverberation mapping [13,14] of BLRs
is one of few methods that can directly derive the masses of
super-massive black holes (SMBHs) in AGNs. In tracing the response of the gas in 
BLRs to the
variable ionizing continuum of AGNs, the time delay between the variations in
the continuum and the broad emission lines gives a characteristic radius of the
BLR gas. With the orbital velocity estimate based on the width
of the broad emission line, the black hole masses can be derived as:
\begin{equation}
M_{BH}=f{{c\tau V^2}\over {G}}
\end{equation}
The scaling factor $f$ is determined by the geometry and velocity distribution
of the BLR gas.
One of the uncertainties in this technique is that the geometry and velocity distribution of the BLR is unknown, and independent techniques are required to
test and calibrate the derived black hole masses. 
Another technique to weigh SMBHs in AGNs is based on the tight relationship 
between the SMBH mass and the velocity dispersion of the bulge or spheroid [15,16].
The $M_{BH}-\sigma_{\ast}$ relationship was primarily discovered in quiescent 
galaxies; however, recent studies have suggested that active SMBHs have
the same correlation [17,18,19].

In this paper we compare the virial products of the infrared time lags and
the narrow Fe K$\alpha$ widths for 10 type 1 AGNs with the black hole masses 
from other techniques. Such study can put strong constraints on the origin 
of the narrow Fe K$\alpha$ line.

\section{DATA AND RESULTS}
We search for archival {\it Chandra} HETG observations for AGNs, as
{\it Chandra} is the only 
instrument currently capable of resolving narrow iron K$\alpha$ lines.
We only choose the objects with narrow Fe K$\alpha$ lines detected at a 
confidence level $>$ 3 $\sigma$. 
We also restrict our sample with BLR reverberation mapping data 
available. The resulting sample consists of ten Seyfert 1 galaxies.
All the archival {\it Chandra} HETG observations of these AGNs are listed in 
Table 1.
The process of data reduction and spectrum fitting is described in [11].
To improve the measurement of line widths, for sources with more than one HETG
exposure, we fitted multiple spectra simultaneously with invariable line 
center energy and FWHM. 
The fitting results are shown in Table 2.

From literature, we find infrared time lags for six of the sources
(see table 2).
From Fig. 30 in [4], we see a tight correlation between
the infrared lag and the $V$ band luminosity, with a scatter of 0.2 dex. Such
a relationship was adopted to estimate the infrared lags for the remaining four
sources without infrared reverberation mapping observations in our sample. 

Peterson et al. presented black hole masses for 35 AGNs based on
broad emission-line reverberation mapping data [20]. The black hole masses are
derived as
\begin{equation}
M_{B}={\frac{f_Bc\tau \sigma_{line}^2}{G}}, f_B=5.5
\end{equation}
where $f_B$ (the scaling factor for BLR reverberation mapping) is a
zero-point calibration determined
using the $M_{BH}-\sigma_{\ast}$ relationship [19,24,25].

Since the narrow Fe K$\alpha$ line is always modeled by a single Gaussian
component, in this research we calculate the virial products as
\begin{equation}
M_{T}={\frac{f_Tc\tau V_{FWHM}^2}{4G}}
\end{equation}
where $f_T$ is the scaling factor, 
$\tau$ the infrared lag and $V_{FWHM}$
the FWHM of the narrow iron K$\alpha$ line. For a Gaussian, FWHM/$\sigma_{line}=2.355$.
Note although the Fe K$\alpha$ line consists of two components (K$\alpha_1$
and K$\alpha_2$), with the spectral resolution of the HETG, its impact on the
measurement of the line width with a single Gaussian is negligible [26].

As shown in Figure 1, nine of the derived
$M_{T}$s are consistent with $M_{B}$s, but a scaling factor is required.
The solid line is the best-fit line (slope fixed to 1) for all ten sources,
derived using the orthogonal regression program GAUSSFIT
(version 3.55; [27]). The asymmetric statistical errors in the
masses were symmetrized as the mean of the positive and negative errors since
GAUSSFIT can not work with asymmetric errors. 
The best-fit scaling factor $f_T$ is $0.44\pm0.19$. 
While excluding the outlier NGC 5548, the best-fit value is $f_T=0.35\pm0.10$.
However, we note the correlation between $M_{T}$ and $M_{B}$ is insignificant.
The Spearman's correlation test yields a confidence level of only 86.5\%
(rho = 0.528), even after excluding NGC 5548.

In Figure 2, $M_{T}$/$f_T$ is plotted versus $\sigma_{\ast}$ for seven sources
with $\sigma_{\ast}$ available in literature (see Table 2). A best-fit was made
to the filled points
\begin{equation}
log ({\frac {M_{T}}{f_T}}) = \alpha + \beta log ({\frac {\sigma_{\ast}}{200}})
\end{equation}
where we adopt $\alpha$ = $8.13\pm0.06$ and $\beta$ = 4.02
as reported by Tremaine et al. for quiescent galaxies [28].
The best-fit yields a scaling factor $f_T$ of $0.42\pm0.23$.
While excluding NGC 5548, we obtained $f_T=0.29\pm0.11$, and a confidence of 
the correlation of 83.3\% (Spearman's rho = 0.617).

\section{DISCUSSION}
Assuming a torus origin of the narrow Fe K$\alpha$ line, we obtained the 
virial products of the narrow iron K$\alpha$ widths and the infrared time
lags for 10 Seyfert 1 galaxies. We found nine out of ten derived virial masses
were consistent with those
masses measured based on reverberation mapping of optical broad emission lines
[20]; however, a scaling factor $\sim 0.4$ is required.
For seven sources with $\sigma_\ast$ measurements, six of them showed virial 
masses consistent with the M-$\sigma_\ast$ relationship, assuming a similar 
scaling factor $\sim 0.4$.

The most common assumption for converting a virial product to $M_{BH}$ is:
\begin{equation}
M_{BH}={\frac {3rV^2_{FWHM}}{4G}},
\end{equation}
which implicitly assumes an isotropic velocity distribution [29].
The scaling factor we obtained indicates that
the narrow Fe K$\alpha$ core is in average 2.6$^{+0.9}_{-0.4}$ times broader than expected
assuming an isotropic velocity distribution of the torus at the distance
measured by the infrared lags.

One possible explanation is that the narrow iron K$\alpha$ line originates from
smaller radius than the infrared radiation does. 
For instance, the obscuring torus might span a large range of scale; the narrow
iron K$\alpha$ line originates from a smaller scale where the temperature is too 
high and dust does not exist, while the infrared radiation comes from a larger 
scale with lower temperatures. Alternatively, the narrow iron K$\alpha$ line 
might be contaminated by the BLR/disk component, thus causing the line width of the 
torus component to be overestimated. If this is true, we would expect a variation 
of the narrow Fe K$\alpha$ line at time scales $\sim$ 6.8 times smaller than
the infrared lags. We note rapid variations of the narrow Fe K$\alpha$ line
has been reported in NGC 7314 [30] and Mrk 841 [31] but not detected in other sources.

Another possibility is that the velocity distribution for dust torus might
not be virial. Here, we consider the dust torus as a thick disk [32].
The thick disk would be sustained vertically
by a pressure which is most probably provided by turbulence with a
characteristic velocity $H/RV_K$, where $R$ is
the radius from the central black hole at the equator and $V_K$ is the local
Keplerian velocity at $R$. Thus, the observed value of $V_{FWHM}$ is given by
\begin{equation}
V_{FWHM} \approx 2V_K\sqrt{(H/R)^2+sin^2i}
\end{equation}
where $i$ is the inclination of
the equator to our line of sight. The observed mass (virial product without any scaling factor)
is:
\begin{equation}
M_{obs}={\frac {c\tau V_{FWHM}^2}{4G}}
\end{equation}
\begin{equation}
M_{obs}=((H/R)^2+sin^2i){\frac {c\tau V_K^2}{G}}
\end{equation}
\begin{equation}
M_{obs}=((H/R)^2+sin^2i)M_{T}
\end{equation}
Hence, 
\begin{equation}
f_T={\frac {1}{(H/R)^2+sin^2i}}
\end{equation}
The observed ratio of Seyfert 2 to Seyfert 1 galaxies of about 4 [33]
would imply a relative torus thickness of $H/R \approx
1.33$, if the distinction is solely due to the orientation of the torus.
Thus a scaling factor $<$ 1 is plausible under this scheme.
It is also clear from the last equation that $f_T$ is less sensitive to the
inclination for H/R $>$ 1. This suggests that if this model for the narrow Fe 
K$\alpha$ line can be confirmed, the narrow Fe K$\alpha$ line could be used
as a good tool to weigh the central SMBH since it is insensitive to the inclination.

In both Fig. 1 and Fig. 2, the only outlier NGC 5548 shows an relatively
underestimated $M_T$.
Cackett \& Horne analyzed 13 years results of optical spectrophotometric
monitoring of the NGC5548 and gave a
luminosity-dependent model of BLRs to explain the variable width of H$\beta$
lines and its variable time-delayed lags [34]. It is possible
that the size of torus also varies with luminosity in this source.
We note that the narrow iron K$\alpha$ line width varied significantly between
two individual exposures; from $5090^{+2020}_{-2030}$ km s$^{-1}$ in 2000 to
$1690^{+1290}_{-1690}$ km s$^{-1}$ in 2002. The line intensity also decreased
from $3.4^{+1.5}_{-1.1}\times10^{-5}$ photons cm$^{-2}$ s$^{-1}$ to
$2.2^{+0.6}_{-0.7}\times10^{-5}$ photons cm$^{-2}$ s$^{-1}$. Meanwhile
the continuum luminosity (2.0 -- 10.0 keV)
increased from $1.5\times10^{43}$ ergs s$^{-1}$ to $1.9\times10^{43}$
ergs s$^{-1}$. This suggests a variable origin of the
narrow iron K$\alpha$ line. The line width presented in Table 2 is dominated
by the second observation with longer exposure time. If simply taking the
line width from the first exposure, we can obtain a black hole mass four times
higher, well consistent with the best-fit lines in Fig. 1 and 2.
For the remaining sources with multiple available HETG exposures,
no narrow iron K$\alpha$ line variation was found to be significant.

While the correlation between $M_{T}$ and $M_{B}$ is statistically insignificant
(at a confidence level of 87\% excluding NGC 5548), we argue that
this may be
attributed to the large uncertainties in the measurements of $M_{T}$ because
of the large errorbars in the iron K line widths.
In Fig. 1 and 2 we can see that the large scattering of data points from
the best-fit lines are mainly due to the large uncertainties in the
measurements of iron K$\alpha$ line width. The scattering of data points in
Fig. 1 can be measured by the
standard deviation of log($M_T/M_B$), which is 0.50 for ten sources
and 0.38 with NGC 5548 excluded.
This is comparable to the typical 1$\sigma$ uncertainty of log($M_T/f_T$)
which is $\sim$ 0.35.

To verify the reliability of this new technique to weigh SMBH, 
more data points and better measurements of iron K$\alpha$ line  widths
are required. Given the difficulty in obtaining lengthily HETG observations
with the required S/N to constrain the iron line width, we have to await the 
next generation of X-ray observatories that are able to measure the iron K 
line at higher spectral resolutions (i.e. of the order 100 km/s) with
calorimeter-based detectors. Before that, we need future X-ray observations to
verify the origins of the narrow iron K$\alpha$ line by better resolving the
line profile and X-ray reverberation-mapping.

\acknowledgments
The work is supported by The Chinese NSF through NSFC10773010 and NSFC 10825312.
We would like to thank Dr. Wei Zheng for helpful comments and his careful review of 
the manuscript. JXW thanks Dr. Tinggui Wang and Matt Malkan for
discussions. Jiang acknowledges support from the "Chuang Xin" Foundation
operated by the Graduate School of USTC.

\begin{figure}
\epsscale{1.}
\plotone{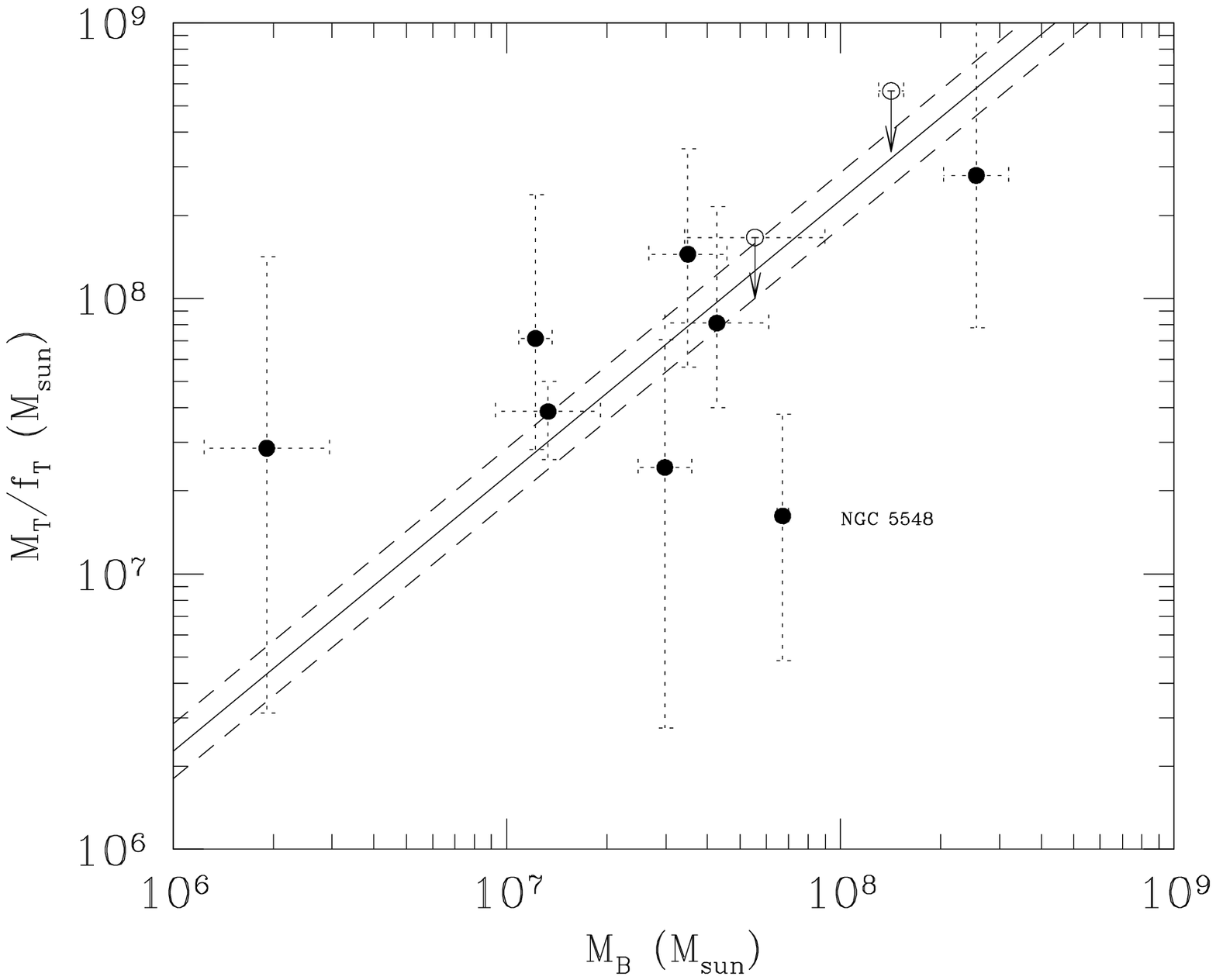}
\caption{$M_{T}$/$f_T$ is plotted versus $M_{B}$. The data are 
presented in Table 2. The solid line is the best-fitting line by fixing the
slope at 1. The dashed lines give the scatter in the normalization.
\label{fig1}}
\end{figure}

\clearpage

\begin{figure}
\epsscale{1.}
\plotone{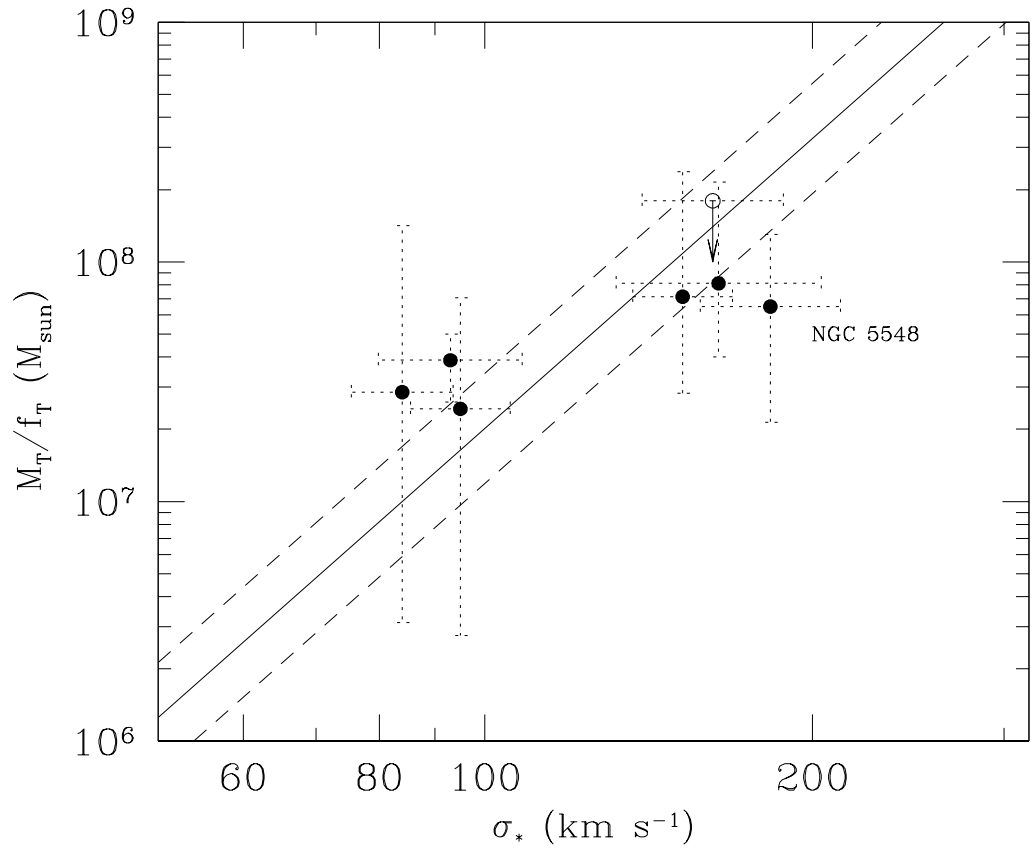}
\caption{$M_{T}$/$f_T$ is plotted versus $\sigma_{\ast}$.
The solid line is the best-fitting line by fixing the slope at 1. The dashed
lines give the scatter in the normalization.
Data are presented in Table 2.
\label{fig2}}
\end{figure}
\clearpage

\begin{deluxetable}{lcc}
\tabletypesize{\normalsize}
\tablecaption{{\it Chandra} HETG Observations \label{tbl-1}}
\tablewidth{0pt}
\tablehead{
\colhead{Object} & \colhead{Start date} & \colhead{Exposure time} \\
\colhead{} & \colhead{} & \colhead{(ksec)}
}
\startdata
NGC 4051 & 2000-03-24 & 80.79 \\
NGC 4151 & 2000-03-05 & 48.03 \\
         & 2002-05-07 & 92.89 \\
         & 2002-05-09 & 156.60 \\
NGC 3783 & 2000-01-20 & 57.16 \\
         & 2001-02-24 & 167.78 \\
         & 2001-02-27 & 171.01 \\
         & 2001-03-10 & 167.57 \\
         & 2001-03-31 & 168.20 \\
         & 2001-06-26 & 168.30 \\
NGC 7469 & 2002-12-12 & 79.89 \\
         & 2002-12-13 & 69.76 \\
Fairall 9 & 2001-09-11 & 79.94 \\
NGC 5548 & 2000-02-05 & 82.32 \\
         & 2002-01-16 & 153.9 \\
NGC 3516 & 2001-04-09 & 36.16 \\
         & 2001-04-10 & 74.54 \\
         & 2001-11-11 & 89.45 \\
MKN 279 & 2002-05-18 & 116.06 \\
MKN 509 & 2001-04-13 & 58.69 \\
3C 120 & 2001-12-21 & 58.16 \\
\enddata
\end{deluxetable}
\clearpage

\begin{deluxetable}{lllllll}
\tabletypesize{\normalsize}
\tablecaption{Estimates of Black Hole Masses of Ten Seyfert 1 Galaxies \label{tbl-2}}
\tablewidth{0pt}
\tablehead{
\colhead{Object} & \colhead{FWHM} & \colhead{$\tau_{IR}$} & \colhead{log ($M_{T}/f_T$)} & \colhead{log $M_{B}$} & \colhead{$\sigma_{\ast}$} & \colhead{References} \\
\colhead{ } & \colhead{(km s$^{-1}$)} & \colhead{(days)} & \colhead{($M_{\sun}$)} & \colhead{($M_{\sun}$)} & \colhead{(km s$^{-1}$)} & \colhead{}
}
\startdata
NGC 4051 & $6160^{+4710}_{-2980}$ & 15.4$\pm$9.1 & $7.456^{+0.695}_{-0.962}$ & 6.281$\pm$0.188 & 84$\pm$9 & 4, 11, 19, 20\\
NGC 4151 & $4070^{+450}_{-630}$ & $48.0^{+2}_{-3}$ & $7.589^{+0.109}_{-0.174}$ & 7.124$\pm$0.157 & 93$\pm$14 & 19, 20, 22\\
NGC 3783 & $2420^{+1590}_{-1580}$ & 85.0$\pm$5 & $7.386^{+0.463}_{-0.945}$ & 7.474$\pm$0.080 & 95$\pm$10 & 11, 19, 20, 23\\
NGC 7469 & $4540^{+2850}_{-1230}$ & 70.9$\pm$18 & $7.854^{+0.521}_{-0.402}$ & 7.086$\pm$0.050 & 152$\pm$16 & 4, 19, 20\\
Fairall 9 & $3780^{+3460}_{-1470}$ & 400.0$\pm$100 & $8.446^{+0.661}_{-0.553}$ & 8.407$\pm$0.097 & & 11, 20, 21\\
NGC 5548 & $2540^{+1170}_{-1080}$ & 51.4$\pm$4.9 & $7.210^{+0.369}_{-0.524}$ & 7.827$\pm$0.017 & 183$\pm$27 & 4, 19, 20\\
NGC 3516 & $4320^{+2700}_{-1290}$ & 89.1 & $7.910^{+0.422}_{-0.308}$ & 7.630$\pm$0.155 & 164$\pm$35 & 11, 19, 20\\
MKN 279  & $5240^{+2890}_{-1970}$ & 109.6 & $8.160^{+0.382}_{-0.410}$ & 7.543$\pm$0.117 & & 11, 20\\
MKN 509  & $<6550$ & 269.2 & $<8.752$ & 8.152$\pm$0.037 & & 1, 6, 7 \\
3C 120 & $<3810$ & 234.4 & $<8.221$ & 7.744$\pm$0.210 & 162$\pm$24 & 11, 19, 20\\
\enddata
\tablecomments{The widths of iron line were reported in the source rest frame. 
Statistical errors for line widths and log($M_T/f_T$) are 90\% confidence.
For the rest quantities, 1$\sigma$ uncertainties are adopted from literature. 
For NGC 4051, NGC 7469 and NGC 5548, we reported the average IR lags observed during the period 2001--2003. For NGC 3516, MKN 279, MKN 509 and 3C 120, the IR lags were estimated based on the relationship $\Delta t \varpropto L^{0.5}$.}
\end{deluxetable}
\end{document}